\documentclass[journal=jacsat,manuscript=article]{achemso}

\usepackage[version=3]{mhchem} 
\usepackage{upgreek}

\author{Li, Zhichao}

\email{0661@pku.edu.cn}

\title[An \textsf{achemso} demo]
  {Two-Nanorod PT Structure with Large Spontaneous Enhancement}

\abbreviations{IR,NMR,UV}

\begin{document}



\begin{abstract}
Two-mode coupling is very common and basic in optics. Recently, a lot more works are focusing on the optical mode coupling with gain and loss due to its tighter link to actual world. For a couple between one gain and one loss mode, by adjusting the coupling constant, the mode in the spectrum will at first stay then move after the constant passing some critical point. This phase transition can be explained by non-Hermitian PT phase transition theory.However, only few works research the spontaneous emission adjustment based on this procedure. It has a lot of promising utilization in on chip optic systems if we can adjust spontaneous emission. We look more deeply in this mode coupling process using metallic nanorods, find the Exceptional Point, show the phase transition between PT symmetry and PT broken regime, and show the adjustment to quantum dot emission. Our work not only promote the understanding of multi-mode system, but also bring up with a new way to adjust high enhancement of spontaneous emission, tamping the foundation for future on chip nano photonics.
 
\end{abstract}

\section{Introduction}
Two-mode coupling is very common and basic in optics.\cite{limonov2017fano} There are works researching two optics mode coupling process, i.e. two nano sphere’s eigen modes coupling, and found out the mode will only shift in a certain region in the parameter space, elsewhere the mode position won’t move.\cite{zhang2006semiconductor} Recent works show that this process can be explained by non-Hermitian PT symmetry phase transition.\cite{guo2009observation} Non-Hermitian system is getting more and more attention because in real experiment losses are unavoidable. The PT symmetric non-Hermitian system brought up in 1990s is causing more attention because it has real spectrum.\cite{bender1998real} Recently, a lot of works are focusing on non-Hermitian PT phase transition in nano photonics system. And in many multi-mode systems with loss Exceptional Points are found.\cite{guo2009observation,ruter2010observation,kodigala2016exceptional,lin2016enhanced,feng2017non,miri2019exceptional,ozdemir2019parity} Such Exceptional Points have a wide range of applications including unidirectional waveguide, sensor, etc.\cite{lin2011unidirectional,chen2017exceptional,park2020symmetry} However, few works are using this phenomena to adjust spontaneous emission enhancement. We focus on the PT phase transition and furthermore discuss its influence to Purcell effect.

When two modes with gain and loss couple to each other, by changing the coupling coefficient the phase shift between PT and PT broken will show up. The phase transfer point is exceptional point, EP.\cite{bender1998real,heiss2004exceptional,heiss2012physics} Similarly, two modes with different losses can also perform this process, such system is called passive PT system. This is very common in optical systems\cite{feng2017non,miri2019exceptional}. In a lot of optical systems, such as two waveguides, nanorods, and optical microcavities, EPs are found.\cite{guo2009observation,ruter2010observation,ding2016emergence,chen2017exceptional,park2020symmetry,ren2021quasinormal} We use two metallic nanorods and let their dipole modes couple to each other to build such PT system, and by changing their distance to adjust the coupling coefficient to achieve PT phase transfer. Besides, with the development of cavity quantum electrodynamics theory,\cite{berman1994cavity,purcell1995spontaneous} metallic nanorods also show a great ability of been a cavity due to their great spontaneous emission enhancement.\cite{vahala2003optical,bohren2008absorption,sauvan2013theory} Some research shows that the enhancement can reach to the level of 10$^4$\cite{kinkhabwala2009large,ridolfo2010quantum,russell2012large,akselrod2014probing}, and by coupling to some wave guide modes, for example, a nanowire, this enhanced spontaneous emission can be guided and put to good use.\cite{lian2015efficient,duan2017large} So we furthermore show the spontaneous emission enhancement of our two-nanorod system, and show how the PT phase transfer will effect this enhancement. 

\section{Theoretical Background}

\begin{figure*}[htpb]
\includegraphics[width=0.8\textwidth]{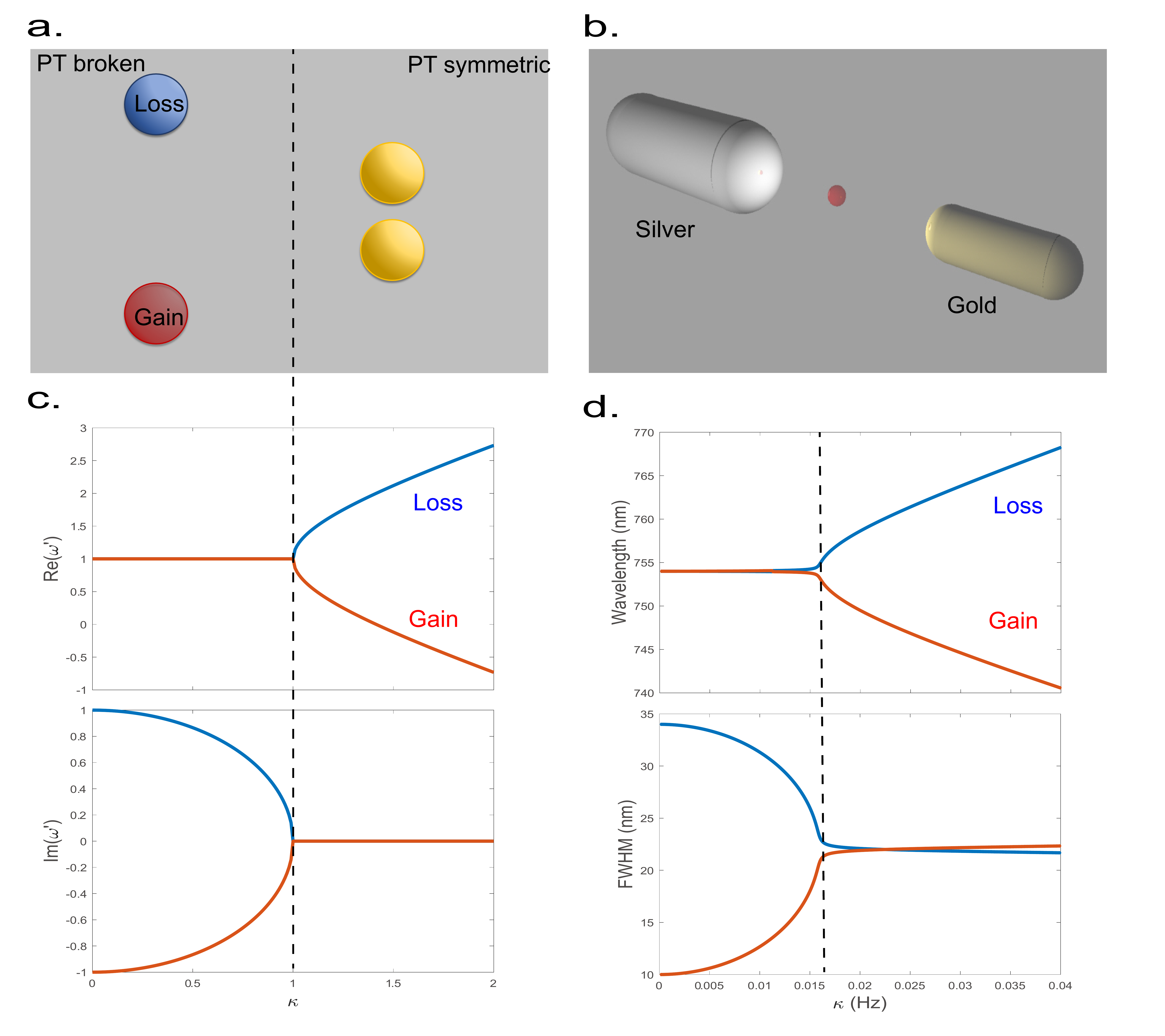}
\caption{\label{fig:epsart} a. The sketch of PT system. In the PT broken phase the energy is localized in the gain mode, while in the PT symmetric phase the energy distributes equally. b. Our two-nanorod PT system, composed by one silver nanorod, one gold nanorod and a emitter at the center. c. The plot of the eigenvalues in equation\ref{eq:2}. d. The plot of the eigenvalues in equation\ref{eq:4}, where the central wavelength of the two original modes are set to be 754nm, and the FWHM to be 34nm and 10nm.}
\label{fig1}
\end{figure*}

First let's focusing on two-mode non-hermitian system. Such kind of system is composed by two modes, one gain and one loss, with the coupling coefficient $\kappa$. We assume the original mode frequency for the two modes are $\omega_1+\gamma_1$, $\omega_2-\gamma_2$, so the system's Hamiltonian can be written as

\begin{equation}
\hat H=\begin{bmatrix}
\omega_1+\gamma_1 & \kappa\\
\kappa & \omega_2-\gamma_1
\end{bmatrix}\label{eq:1}.
\end{equation}

To make this system a PT symmetric system, the PT transform of Hamiltonian should be the same as the original one, means that $\hat P\hat T\hat H=\hat H$. Under that condition, the original frequencies should follow the equation that $\omega_1+\gamma_1=\omega_2+\gamma_2$. So we have $\omega_1=\omega_2=\omega$ and $\gamma_1=\gamma_2=\gamma$. For simplicity we set $\omega=1$,$\gamma=1$, then find out the eigenvalue for this Hamiltonian to be
\begin{equation}
\left\{
\begin{aligned}
\omega'_1=1+\sqrt{\kappa^2-1}\\
\omega'_2=1-\sqrt{\kappa^2-1}
\end{aligned}
\label{eq:2}
\right.
\end{equation}
  Assume that the $\kappa$ is zero or a positive real number. It's easy to find out that when $0<\kappa<1$ the square root number is negative, so the two eigenvalues will have the same real part and opposite imaginary part. This is in the PT broken phase. When $\kappa>1$, the square root number is positive, the two eigenvalues will have same imaginary part but real part symmetric to 1. This is in the PT phase. When $\kappa=1$, the two eigenvalues are the same and this Hamiltonian only has one eigenvector. This $\kappa=1$ is the EP. 

In the PT broken phase, since the two eigen modes have opposite imaginary part, there will be one gain mode and one loss mode, and the field will mostly localized in the gain mode. While in the PT broken mode, the imaginary part of eigenvalues is 0, so the field will equally distributed in both eigen nodes. These two kinds of field are showed qualitatively in FIG\ref{fig1}a. And the corresponding eigenvalue behavior is showed in FIG\ref{fig1}c. Moreover, this behaviour won't change when we add a constant imaginary part to both original mode frequencies, meaning that the two original modes don't have to have one gain mode. Two modes with different loss can also generate the PT system. Since there are no gain in the system, it's a passive PT system.

Now we try to use two different metallic nanorods to build such system. The structure is showed in FIG\ref{fig1}b. Two nanorods are placed co-axially. They have different material and different size, but share the same dipole mode wavelength. By changing their distance we can control their coupling coefficient. According to the coupled mode theory,\cite{awai2006overlap,elnaggar2015energy}, we give the theoretical equation for two metallic nanorods' coupling.

\begin{equation}
\left\{
\begin{aligned}
\ddot a_1+\tilde\omega_1^2a_1+\tilde\omega_2^2\kappa a_2=0 \\
\ddot a_2+\tilde\omega_2^2a_2+\tilde\omega_1^2\kappa a_1=0
\end{aligned}
\label{eq:3}
\right.
\end{equation}

Here $a_1$, $a_2$ represent the amplitude of two modes, $\tilde\omega_1$, $\tilde\omega_2$ are two complex numbers, their real part show the eigen frequency and the imaginary part show the loss. And $\kappa$ shows the coupling between the two original modes. By solving the differential equation(\ref{eq:3}), we get the expression of eigen frequencies $\tilde\omega_1'$, $\tilde\omega_2'$ after coupling.

\begin{equation}
\left\{
\begin{aligned}
\tilde\omega_1'^2=\frac{1}{2}(\tilde\omega_1^2+\tilde\omega_2^2)+\sqrt{(\frac{\tilde\omega_1^2-\tilde\omega_2^2}{2})^2+\tilde\omega_1^2\tilde\omega_2^2\kappa^2} \\
\tilde\omega_1'^2=\frac{1}{2}(\tilde\omega_1^2+\tilde\omega_2^2)-\sqrt{(\frac{\tilde\omega_1^2-\tilde\omega_2^2}{2})^2+\tilde\omega_1^2\tilde\omega_2^2\kappa^2} \\
\end{aligned}
\label{eq:4}
\right.
\end{equation}

By changing $\kappa$ in equation(\ref{eq:4}), the PT phase transfer is clear to see. We make a plot based on equation(\ref{eq:4}) in FIG\ref{fig1}d. The results fit the simple model of PT system. Next we use simulation to show the PT phase transfer in our two nanorod system.

\section{Model and Simulation}

\begin{figure*}[htpb]
\includegraphics[width=0.8\textwidth]{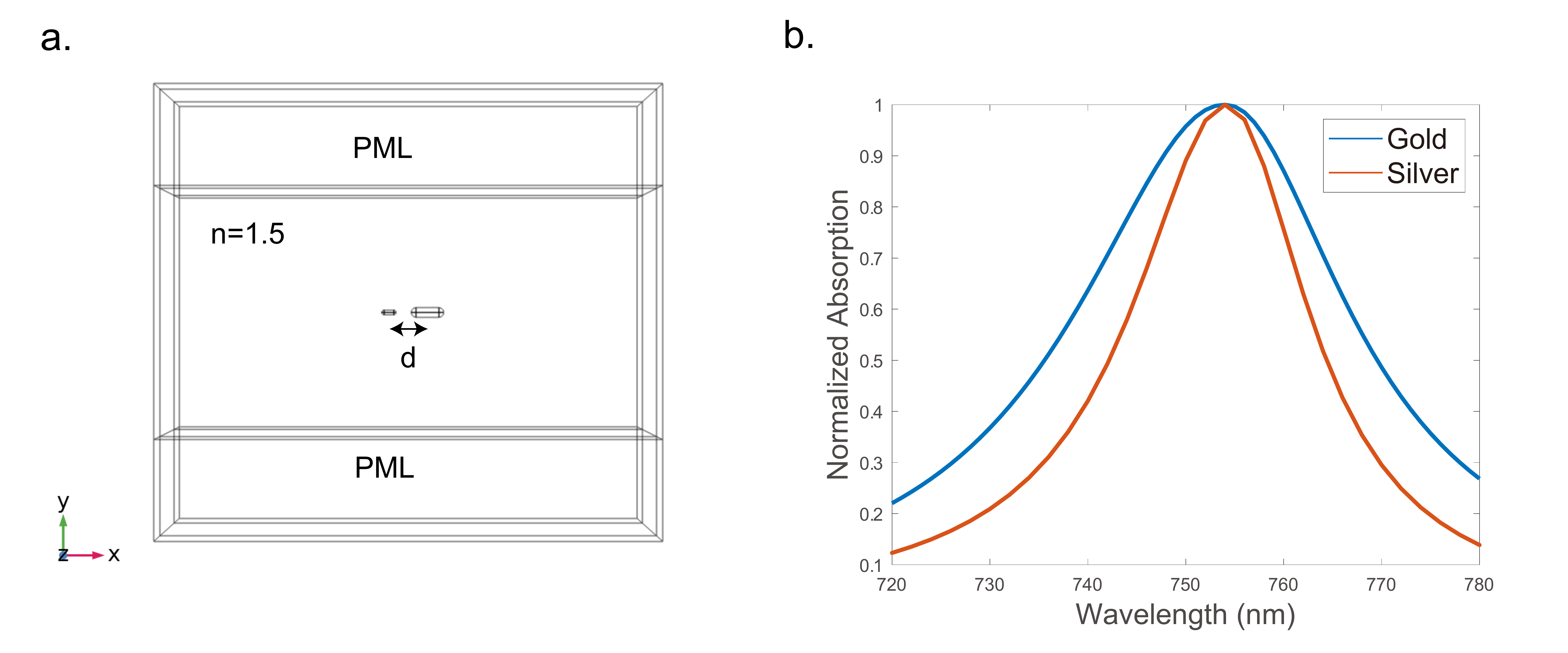}
\caption{\label{fig:epsart} a. The Comsol model for our simulation. The background has the optical constant n=1.5. The X min and X max boundaries are set to be periodic boundary condition. The other boundaries are coated with PML. The two nanorods are placed with the center-to-center distance $d$. The plane wave light source is polarized in X direction. b. The absorption spectrum for single gold and silver nanorod. They both have the absorption peak at 754nm, and the gold nanorod's peak FWHM is 34nm, and the silver nanorod's peak FWHM is 10nm.}
\label{fig2}
\end{figure*}

We use Comsol to build our model and simulate, the model is showed in FIG\ref{fig2}a. We use two semi spheres and a cylinder to build a nanorod, and we put one silver and a gold nanorod co-axially. The optical constants are from \cite{johnson1972optical}. The whole structure is set in the environment with the constant refractive index 1.5. In the axial direction we choose periodic condition with the period of 1000nm, and other boundary we choose PML condition. We use plane wave source to excite the structure, the polarization of the source is along the axis direction.

According to the derivation, we need to make the two original modes the same frequency and different loss. So the nanorod modes should have the same wavelength and different FWHM. We choose the radius of the gold nanorod to be 5nm, the height of the cylinder of gold nanorod to be 20nm. For the silver nanorod, the radius is 10nm and the height is 47.55nm. We do the simulation with a single nanorod to see the absorption spectrum. The plot is showed in FIG\ref{fig2}b. This design allows the two 
nanorod modes to have the same wavelength of 754nm. And the FWHM for gold nanorod is 34nm, for silver is 10nm. This is the base to find the PT phase transition and the EP.

\section{Result and Discussion}

\begin{figure*}[htpb]
\includegraphics[width=0.8\textwidth]{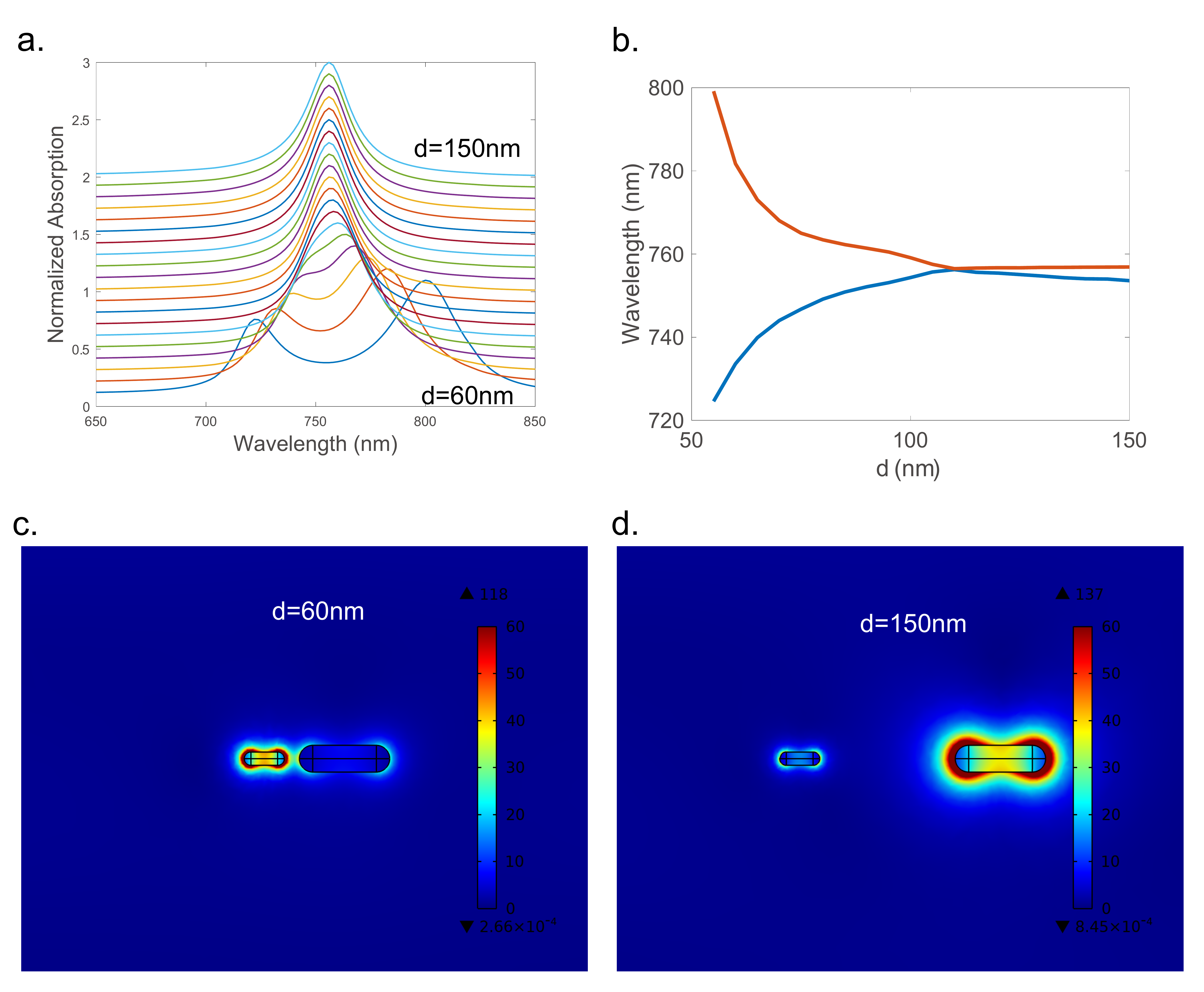}
\caption{\label{fig:epsart} a. The absorption spectrum of our structure with $d$=60nm-150nm, with the step of 5nm. b. The two Lorentzian fitting results of spectrum in a. The cross point of two fitted wavelength lines is $d$=110nm, which means that the EP in our structure is around $d_{EP}$=110nm. c. The electric field profile for $d$=60nm at 754nm. d.The electric field profile for $d$=15nm at 754nm.  }
\label{fig3}
\end{figure*}

Then we simulate the two-nanorod structure. By changing the center-to-center distance between the two nanorods $d$, from 60nm to 150nm, we adjust the coupling coefficient in this system. The total absorption spectrum of each distance is showed in FIG\ref{fig3}a. It's clear that when the distance decreasing from 150nm, before it passes the EP, the change of the distance won't effect the single peak position. But when the distance goes below the EP, the single peak will split into two peaks and when the distance continuously decreasing the split is getting larger. This feature shows the PT phase transition. To precisely determine the EP, we fit the spectrum using double Lorentzian peak and get the frequency and the loss. According to FIG\ref{fig3}b. the EP in our system is around $d_{EP}$=110nm.

Moreover, we show the field profile in both single-peak phase and double-peak phase to prove that we actually see the PT transition. The field profile of $d$=60nm and $d$=150nm are showed in FIG\ref{fig3}c. and FIG\ref{fig3}d. In the PT symmetric phase, $d$=60nm, the electric field in the hot spot of both nanorods are basically the same, while in the PT broken phase, $d$=150nm, the electric field is larger in the silver nanorod than in the gold one, means that the field is more localized in the lossless mode. This feature also shows that we find the PT phase transition.

\begin{figure*}[htpb]
\includegraphics[width=0.8\textwidth]{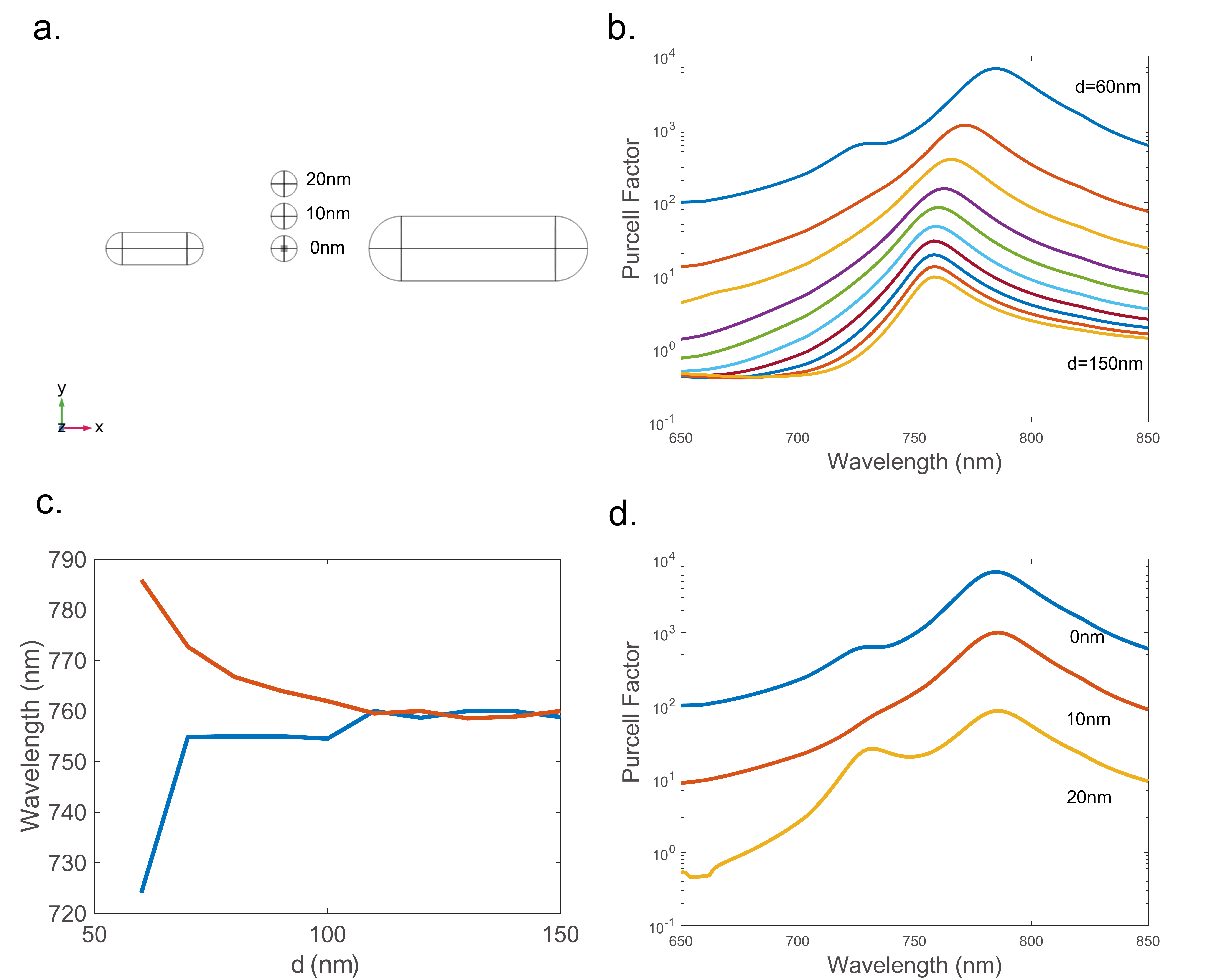}
\caption{\label{fig:epsart} a. Changing the light source to emitter. We put a point dipole source in the center of sphere. We also put the source off the axis to see the change. The off-axis distance changes from 0 to 20nm. b. The Purcell factor spectrum of our structure with $d$=60nm-150nm, with the step of 10nm. c. The two Lorentzian fitting results of spectrum in b. The cross point of two fitted wavelength lines is also $d$=110nm. d.The Purcell factor spectrum for $d$=60nm with the off-axis distance equals to 0,10 and 20nm. The Purcell factor decreases with the distance increase, but the peak wavelength doesn't change.}
\label{fig4}
\end{figure*}

Based on the simulation result above, we are testing the spontaneous emission enhancement of our structure. We change the source to be a dipole emitter to simulate a quantum dot, and put it at the center of the two nanorods' gap, showed in FIG\ref{fig4}a. And we change the position of both nanorods spontaneously to keep the emitter the same place. We simulate the total absorption spectrum and calculate the Purcell effect of our structure. The spectrum is showed in FIG\ref{fig4}b.

We can still see the PT phase change that when shrinking the gap between nanorods, the single peak splits to two. Besides, we also see that when the gap shrinks, the total Purcell factor increases meaning the spontaneous emission enhancement is increasing. At the same time, we find that although the amplitudes of the two eigenmodes should be equal in the PT symmetric phase, the Purcell factor spectrum is still more localized in the gold nanorod, while the proportion of localization is lower than that in the PT broken phase. We also fit the obtained Purcell spectral line, and the fitting result showed in FIG\ref{fig4}c. This further verified that our system reaches the EP near $d_{EP}$=110nm. We also show the results when the emitter is not perfectly at the central axis. The result in FIG\ref{fig4}d. shows that when the emitter is further away from the central axis, the Purcell factor is smaller, while the peak won't shift.

\section{Conclusion}

We find the phenomenon of PT phase transition in two nanorod coupled systems and find the EP. We verify the existence of PT phase transition from two aspects of the eigenvalue curve and electric field distribution. In addition, we find that this system not only has a large Purcell coefficient, but also the center wavelength of spontaneous emission enhancement can be shifted due to the PT phase transition. This also lays a foundation for the subsequent study of the tunable structure of spontaneous emission enhancement.

\section{Acknowledgement}

Thank the help of professor Ying Gu and her students Ma Yun, Zhang Xinchen, et.al from State Key Laboratory for Mesoscopic Physics, Frontiers Science Center for Nano-optoelectronics, Collaborative Innovation Center of Quantum Matter, and Beijing Academy of Quantum Information Sciences, Department of Physics, Peking University.

\nocite{*}

\bibliography{apssamp}

\end{document}